\documentstyle[here,epsfig]{mn2e}
\def\simlt{\mathrel{\rlap{\lower 3pt\hbox{$\sim$}}\raise 2.0pt\hbox{$<$}}}
\def\simgt{\mathrel{\rlap{\lower 3pt\hbox{$\sim$}} \raise 2.0pt\hbox{$>$}}}

\def\gtsima{$\; \buildrel > \over \sim \;$}
\def\ltsima{$\; \buildrel < \over \sim \;$}
\def\gtrsim{\lower.5ex\hbox{\gtsima}}
\def\lesssim{\lower.5ex\hbox{\ltsima}}
\def\url#1{{\ttfamily\def\/{/\diskretionary{}{}{}}#1}}
\begin{document}

\newcommand{\q}{\begin{equation}}
\newcommand{\qa}{\begin{eqnarray}}
\newcommand{\qs}{\begin{eqnarray*}}
\newcommand{\nq}{\end{equation}}
\newcommand{\nqa}{\end{eqnarray}}
\newcommand{\nqs}{\end{eqnarray*}}
\newcommand{\ud}{\mathrm{d}}

\title[Constraints on IMBHs] 
{Constraints on Galactic Intermediate Mass Black Holes}
\author[M. Mapelli, A. Ferrara \& N. Rea] 
{M. Mapelli$^{1}$,  A. Ferrara$^{1}$ \& N. Rea$^{2}$ \\
$^{1}$SISSA, International School for Advanced Studies, Via Beirut 4, 34100, Trieste, Italy; {\tt mapelli@sissa.it}\\
$^{2}$SRON, Netherlands Institute for Space Research, Sorbonnelaan, 2, 3584 CA, Utrecht, The Netherlands\\}

\maketitle \vspace {7cm }

\begin{abstract}
Intermediate Mass Black Holes (IMBHs; 10$^{1.3-5}\,{}M_\odot$) are thought to form as relics of Population III stars 
or from the runaway collapse of stars in young clusters; their number and very existence are uncertain. 
We ran N-body simulations of Galactic IMBHs, modelling them as a halo population distributed according to 
a Navarro, Frenk \& White (NFW) or a more concentrated Diemand, Madau \& Moore (DMM) density profile. 
As IMBHs pass through Galactic molecular/atomic hydrogen regions, they accrete gas, 
thus becoming X-ray sources. We constrain the density of Galactic IMBHs, $\Omega_\bullet$, by comparing the 
distribution of simulated X-ray sources with the observed one.  From the null detections of Milky Way Ultra-Luminous 
X-ray sources, and from a comparison of simulations with unidentified sources in the IBIS/ISGRI catalogue we find a strong 
upper limit $\Omega_\bullet\le 10^{-2}\Omega_b (\le 10^{-1}\Omega_b)$ for a DMM (NFW) profile, if IMBHs accrete via 
ADAF disks. Slightly stronger constraints ($\Omega_\bullet \le {} 10^{-3}\Omega_b$ for a DMM profile; $\Omega_\bullet\ 
\le 10^{-2}\Omega_b$ for a NFW profile) can be derived if IMBHs accrete with higher efficiency, such as by forming 
thin accretion disks. Although not very tight, such constraints are the most stringent ones derived so far in the 
literature. 
\end{abstract}

\section{Introduction}
%Part I: literature\\
There are strong observational evidences for the existence of two classes of black holes (BHs): stellar mass BHs, with mass from 3 to 20 $M_\odot{}$  (Orosz 2003), thought to be the relics of massive stars, and super massive black holes (SMBHs) in the mass range
$10^{6-9} M_\odot$, hosted in the nuclei of many galaxies. Recently, the existence of a third class of  Intermediate Mass BHs (IMBHs)
has been inferred. They are
characterized by masses in the range from $20 M_\odot{}$ to a few $\times 10^4M_\odot$ (see van der Marel 2004 for a review).
 
Several IMBH formation mechanisms have been proposed:  {\it (i)} IMBHs could be the relics of very massive metal free stars (Heger \& Woosley 2002), {\it (ii)} they could form in young clusters via runaway collapse of stars (Portegies Zwart \& McMillan 2002), or {\it (iii)} could have been built up in globular clusters through repeated mergers of stellar mass BHs in binaries (Miller \& Hamilton 2002). Some recent observations indicate that IMBHs could exist in the core of globular clusters (Gebhardt, Rich \& Ho 2002, 2005; Gerssen et al. 2002; van den Bosch et al. 2005). Their number is nearly unknown. In principle, IMBHs could contribute to all the baryonic dark matter (van der Marel 2004): their  density $\Omega_\bullet $ could be essentially equal to that of luminous baryonic matter,  $\Omega_{b,\,{}lum}=0.021$ (Persic \& Salucci 1992; Fukugita, Hogan \& Peebles 1998), and equal to  50\% of all baryons ($\Omega_b=0.044$,  Spergel et al. 2003) . Only weak constraints on the IMBH mass have been inferred from dynamical studies of the Milky Way. For example, the observed velocity dispersion of stars in the Galactic disk requires that halo BHs masses are $\le 3\times 10^6M_\odot$, if the Milky Way dark halo is entirely made of  compact objects (Carr \& Sakellariadou 1999; see also Lacey \& Ostriker 1985; Wasserman \& Salpeter 1994; Murali, Arras \& Wasserman 2000 and reference herein). Other dynamical constraints on the IMBH mass can be derived by imposing that they do not disrupt too many Galactic globular clusters (Moore 1993; Arras \& Wasserman 1999). By this request, Klessen \& Burkert (1996) found that, if the dark halo of the Milky Way is exclusively composed by IMBHs, their mass must be  $\le 5\times{}10^4\,{}M_\odot{}$. For the same reason, the halo BHs cannot represent more than 2.5-5 per cent of the dark halo mass, if they are as massive as $10^6M_\odot{}$ (Murali et al. 2000). However, constraints on IMBHs obtained from globular cluster disruption are very uncertain, as we do not know what are the characteristics of globular clusters when they form, and how many of them have been destroyed. It could even be that IMBHs  have played a role in determining the current number and distribution of globular clusters (Ostriker, Binney \& Saha 1989).

In this paper we explore an alternative way to dynamically constrain the IMBH density, based on the proposed identification of Ultra-Luminous X-ray sources (ULXs) with IMBHs. ULXs are, by definition, point sources with X-ray luminosity higher than $10^{39}$ erg s$^{-1}$, exceeding the isotropic Eddington limit for a 10 $M_\odot{}$ BH (see Colbert \& Miller 2005 for a review). ULXs have not been found, up to now, in the Milky Way; but they are present  in many spiral and starburst galaxies (Swartz et al. 2004). 
ULXs tend to be associated with star forming regions; but they often lie near, not in them (Mushotzky 2004).

ULXs were initially identified with SMBHs  with a low accretion rate; but this interpretation was found to be in conflict with their position in the host galaxies, far off from the galaxy center (Colbert \& Miller 2005). Later on ULXs have been suggested to be high-mass X-ray binaries (HMXBs) with beamed X-ray emission (King et al. 2001; Grimm, Gilfanov \& Sunyaev 2003; King 2003). Even though low luminosity ULXs ($L_X\lesssim{}5\times{}10^{39}$ erg s$^{-1}$) are consistent with this HMXB scenario, the highest luminosity ULXs show various characteristics which can be hardly reconciled with the beaming model (Miller, Fabian \& Miller 2004), such as the existence of a ionized nebula surrounding some bright ULXs (Pakull \& Mirioni 2002; Kaaret, Ward \& Zezas 2004). 

An intriguing hypothesis, at least for these highest luminosity ULXs (Miller et al. 2004), is their identification with accreting IMBHs. This idea is also supported by some observational evidences. First, the spectra of many high luminosity ULXs have a soft component well fitted by a multicolor black-body disk, whose inner temperature is typical of  BH masses in the IMBH range (Miller et al. 2004; Colbert \& Miller 2005). In addition, high luminosity ULXs often show long term variability on timescales of months to years (Kaaret et al. 2001; Matsumoto et al. 2001; Miyaji, Lehmann \& Hasinger 2001) and quasi periodic oscillations (Strohmayer \& Mushotzky 2003), inconsistent with the beaming scenario. King \& Dehnen (2005) propose that very high luminosity ULXs in interacting galaxies can be IMBHs hosted in the merging satellite and whose accretion is activated by tidal forces.
%in a satellite which is merging into a larger halo. The tidal forces due to the merger process can push many stars of the satellite on to radial orbits, feeding the IMBH.

Many studies have been dedicated to check the possibility that ULXs are IMBHs accreting in binary systems (Baumgardt et al. 2004; Hopman, Portegies Zwart \& Alexander 2004; Kalogera et al. 2004; Portegies Zwart, Dewi \& Maccarone 2004; Hopman \& Portegies Zwart 2005; Patruno et al. 2005). However, the observed population of ULXs is not well reproduced by this binary system scenario (Blecha et al. 2005; Madhusudhan et al. 2005), mainly because the ULX phase of simulated IMBHs is too short. A better agreement between simulations and observations can be obtained only by considering very massive IMBHs ($\gtrsim{}1000\,{}M_\odot{}$; Baumgardt et al. 2005). In addition, very few optical counterparts have been detected  for ULXs so far and can unambiguously be identified as companion stars (Liu et al. 2005; Colbert \& Miller 2005). Then, it is still open the possibility that ULXs are IMBHs accreting gas during the transit through a dense molecular cloud, as recently suggested by Krolik (2004) and by Mii \& Totani (2005). 

This paper is aimed at exploring in detail this last hypothesis by dedicated N-body simulations (Section 2).
In particular, we simulate a Milky Way model and we derive an upper limit of the density of IMBHs, by requiring that no ULX is produced in the Milky Way by IMBHs passing through molecular clouds (Section 3).
Next, we study the distribution of both ULX and non-ULX sources produced by IMBHs passing through  molecular clouds (Section 3) and atomic hydrogen regions (Section 4). We finally compare the derived distributions with observations (Section 5).

%\section{METHOD}
\section{Numerical simulations}
The simulations have been carried out using the parallel N-body code Gadget-2 (Springel 2005). The simulations were performed using 8 nodes of the 128 processor cluster {\it Avogadro} at the {\it Cilea} ({\url http://www.cilea.it}).  Our aim is to generate a suitable N-body model of the Milky Way, in which we embed a halo population of IMBHs.

\subsection{Milky Way model}
To reproduce the Milky Way we simulated an exponential disk and a Hernquist spherical bulge, whose density profiles are given, respectively, by the following relations (Hernquist 1993):
\q\label{eq:eq1}
\rho_d(R,z)=\frac{M_d}{4\pi R_d^2\,{}z_0}\,{}\exp{-(R/R_d)}\,{}\textrm{sech}^2(z/z_0)
\nq
\q\label{eq:eq2}
\rho_b(r)=\frac{M_b\,{}a}{2\pi}\frac{1}{r\,{}(a+r)^3},
\nq
where $M_d$ ($M_b$) is the disk (bulge) mass, $R_d$ is the disk scale radius, $z_0$ is the disk scale height, $a$ is the bulge scale length and $r=\sqrt{R^2+z^2}$. We choose $a=0.2\,{}R_d$, consistent with Kent, Dame \& Fazio (1991; $a=0.7\pm{}0.2$ kpc). The value of $z_0$ is quite difficult to constrain. Recent observations (Larsen \& Humphreys 2003; Yoachim \& Dalcanton 2005) suggest the presence of two components in the thin disk of the Milky Way: a ''young star forming'' thin disk (with scale height $\sim{}200$ pc) and  an ''old'' thin disk (with scale height $\sim{}600$ pc). Then, we assume $z_0=0.1\,{}R_d=350$ pc, which is approximately an average of the scale height of these two components and correlates in a simple way with $R_d$.

Disk and bulge are embedded in a rigid dark matter halo, whose density
profile is (Navarro, Frenk \& White 1996, hereafter NFW; Moore et al. 1999):
\q\label{eq:eq3}
\rho_h(r)=\frac{\rho_s}{(r/r_s)^\gamma{}\,{}[1+(r/r_s)^\alpha{}]^{(\beta{}-\gamma{})/\alpha{}}},
\nq
where we choose $(\alpha{},\beta{}, \gamma{})=(1,3,1)$, and
$\rho{}_s=\rho_{crit}\,{}\delta_c$, $\rho_{crit}$ being the critical
density of the Universe and
\begin{equation}\label{eq:eq4}
\delta_c=\frac{200}{3}\frac{c^3}{\ln{(1+c)}-(c/(1+c))},  
\end{equation}
where $c$ is the concentration parameter and $r_s$ is the halo scale radius, defined by
$r_s=R_{200}/c$; $R_{200}$ is the radius encompassing a mean overdensity of 200 with respect to the 
background density of the Universe, i.e. the radius containing the virial mass $M_{200}$. Given the concentration $c$ and
the Hubble parameter\footnote{We adopt $H(z\lesssim{}1)=H_0=71$ km s$^{-1}$
Mpc$^{-1}$ in agreement with first year WMAP results (Spergel et
al. 2003)} $H(z)$, $R_{200}$, $M_{200}$ and the circular velocity at the
virial radius, $V_{200}$, are related by the following expressions.
\qa\label{eq:eq5}
%\begin{array}{l}
R_{200}=\frac{V_{200}}{10\,{}H(z)}\hspace{0.5cm}\nonumber{}\\
M_{200}=\frac{V_{200}^3}{10\,{}G\,{}H(z)};
%\end{array}
\nqa
$G$ is the gravitational constant. In our simulations we fix $c=12$, $V_{200}=160$ km s$^{-1}$ (see
Klypin, Zhao \& Somerville 2002), yielding $M_{200}=1.34\times{}10^{12}M_\odot{}$, $R_{200}=225$ kpc and $r_s=19$
kpc (Table 1 reports the initial parameters); finally, we use for $H(z)$ its actual value $H_0$. 

Rigid halos can induce $m=1$ instabilities in the disk. To check this effect, we performed test simulations with a non-rigid halo (with halo particles ten times more massive than disk particles); we did not observe significant differences in the evolution with respect to fixed-halo models. Since simulations with non-rigid halos are prohibitively time consuming for the  very high resolutions required by the problem (see Section 2.3 for details), we have chosen to adopt a rigid halo. 

To derive $M_d$, $M_b$ and $R_d$ we followed the prescriptions of Mo, Mao \& White (1998), imposing that the disk is a thin, dynamically stable and centrifugally supported structure, whose mass is a fraction of the halo mass and whose angular momentum is a fraction of the halo
angular momentum. In particular, our best, stable model is obtained
for a choice of the spin parameter $\lambda{}=0.035$, where
$\lambda{}\equiv{}J\,{}|E|^{1/2}G^{-1}M^{-5/2}$ ($J$, $E$ and $M$
being the angular momentum, the total energy and the mass of the
halo, respectively). Requiring that $M_d+M_b\approx{}0.04\,{}M$ and that $M_d:M_b=4:1$
(in agreement with Kent et al. 1991; Freudenreich 1998; Binney \&
Merrifield 1998), we obtain $M_d=4\times{}10^{10}\,{}M_\odot{}$ and
$M_b=1\times{}10^{10}\,{}M_\odot{}$. Our choices are in agreement both
with the best model of Milky Way described in Klypin et al. (2002) and with the COBE measurements of the bulge mass
($M_b=1.3\pm{}0.5\times{}10^{10}M_\odot{}$, Dwek et al. 1995). For
these values, the scale radius of the disk is $R_d=3.5$ kpc,
consistent with recent estimates (Binney \& Merrifield 1998).  Given
the uncertainty on the $M_d/M_b$ ratio, we also made some test
simulations for $M_d:M_b=5:1$ observing no significant differences in
our results. In order to account for the SMBH in the nucleus of
the Milky Way, we located a point mass $M_{SMBH}=3.5\times{}10^{6}M_\odot{}$ (Ghez et al. 2003;
Sh$\ddot{\textrm{o}}$del et al. 2003) at the center of the rigid halo.

Initial velocities of disk and bulge particles are simulated using the
Gaussian approximation (Hernquist 1993) for dispersion
velocities. This choice introduces a transient behavior, represented
by outwards propagating rings of overdensity from the warmer disk
center, as it was already noted by Kuijken \& Dubinsky 1995 (see
also Kazantzidis, Magorrian \& Moore 2004; Widrow \& Dubinski 2005).
In agreement with the findings of Kuijken \& Dubinsky 1995, in the
highest resolution runs (when the mass of each particle is
$m\lesssim{}10^5M_\odot{}$ and the total number of particles
approaches one million) this transient is stronger; nevertheless, it always
disappears within about 1 timescale\footnote{The timescale of our
simulation is defined as the rotation period of the simulated galaxy,
i.e. about 0.27 Gyr.}, when the system relaxes into a new equilibrium
configuration. We consider this new relaxed configuration as initial
condition for our analysis. This procedure is legitimate in our case,
since we are not investigating processes such as disk instabilities,
but we are only interested in the dynamical evolution of halo
IMBHs. \\ After relaxation, we continue the simulation for
about 5 Gyr, i.e. from redshift $z\approx 0.5$ until today, 
 about half of the time elapsed from the last major merger 
(Governato et al. 2004). This allows us to follow the evolution of an already
relaxed and nearly unperturbed (by mergers) Milky Way.  During the
entire simulation  disk and bulge remain perfectly stable.

%%%%%%%%%%%%%%%%%%%%%%%%%%%%%%%%%%% TABLE 1 %%%%%%%%%%%%%%%%%%%%%%%%%%%%%%%%%%%
\begin{table}
\begin{center}
\caption{Initial parameters for the Milky Way model.
}
\begin{tabular}{ll}
\hline
\hline
\vspace{0.1cm}
$c$ & 12\\
$V_{200}$ & 160 km s$^{-1}$\\
$M_{200}$ & 1.34$\times{}10^{12}M_{\odot{}}$\\
$R_{200}$ & 225 kpc \\
$\lambda{}$ & 0.035\\%0.04
$M_d/M_b$ & 4\\
$M_d$ & 4$\times{}10^{10}M_\odot{}$\\
$M_b$ & 10$^{10}M_\odot{}$\\
$R_d$ & 3.5 kpc\\
$z_0$ & 0.1 $R_d$\\
$a$ & 0.2 $R_d$\\
\hline
\end{tabular}
\end{center}
%\begin{flushleft}
%{\footnotesize $^{a}$Totani \& Takeuchi model rescaled to the Spitzer data.}\\
%{\footnotesize $^{b}$(K) and (W) indicate the ZL subtraction obtained using Kelsall's model and Wright's model, respectively.}\\
%\end{flushleft}
\label{tab_1}
\end{table}
%%%%%%%%%%%%%%%%%%%%%%%%%%%%%%%%%%%%%%%%%%%%%%%%%%%%%%%%%%%%%%%%%%%%%%%%%%%%%%%

\subsection{Intermediate mass black holes}
How many IMBHs are hosted in the Milky Way ? What is their spatial
distribution ? These are yet unanswered questions.
Nevertheless, we need an $Ansatz$ on the IMBH number, mass and
distribution to generate the initial conditions of our simulations.
A reasonable estimate for the initial IMBH number follows from
Volonteri, Haardt \& Madau 2003. Assuming that the IMBHs are born in
$\nu\,{}\sigma$ fluctuations (with $\nu=3-3.5$) collapsing at a
given redshift, Volonteri et al. (2003) derive the density of IMBHs at
the formation redshift $\Omega{}_{\bullet{},\,{}f}$ as
\q\label{eq:eq6}
\Omega_{\bullet,f}=\left[1-\textrm{erf}\left(\nu/\sqrt{2}\right)\right]\,{}\Omega_M\,{}\frac{m_{\bullet,f}}{M(\nu{})},
\nq
where $[1-\textrm{erf}(\nu/\sqrt{2})]\Omega_M$ is the fraction of the Universe matter in halos with $M>M(\nu{})$ ($\Omega_M=0.27$ being the matter density), as derived from the Press \& Schechter (1974) formalism, and $m_{\bullet,f}/M(\nu)$ is the fraction of mass of the halo collapsed in IMBHs ($m_{\bullet{},f}$ being the average initial IMBH mass, and $M(\nu{})$ the mass of the $\nu\,{}\sigma{}$ peak halo). For example, assuming that IMBHs form in $3\,{}\sigma{}$ fluctuations collapsing at redshift $z=24$, the corresponding halo mass is $M(3)=1.7\times{}10^5M_\odot{}$ (Barkana \& Loeb 2001). Under these assumptions equation (\ref{eq:eq6}) gives $\Omega_{\bullet,f}=10^{-4}\,{}\Omega_b\,{}(m_{\bullet{},f}/10^3M_\odot)$.

Given $\Omega_{\bullet{},\,{}f}$, one can roughly estimate the number of IMBHs in the Milky way, $N_{\bullet{}}$ as
\q\label{eq:eq7}
N_{\bullet{}}=\frac{\Omega_{\bullet{},\,{}f}}{\Omega{}_b}\frac{M_{b,MW}}{m_{\bullet{},f}},
\nq
where $M_{b,MW}=(0.5-1)\times{}10^{11}M_\odot{}$ is the mass in baryons of the Milky Way. Adopting a value of $\Omega_{\bullet{},\,{}f}=10^{-4}\,{}\Omega_b$, we find $N_{\bullet{}}\approx{}10^4$. Instead, if we assume that IMBHs form in $3.5\,{}\sigma{}$ fluctuations collapsing at $z=24$, this number becomes $N_{\bullet{}}=5\times{}10^2$. We will adopt equations (\ref{eq:eq6}) and (\ref{eq:eq7}) to calculate how many IMBHs to include in our simulations.

How massive are the IMBHs today? We have assumed that their average mass at formation was $m_{\bullet{},f}=10^3\,{}M_\odot{}$. However, it is likely that they accreted gas for some period of their life (Ricotti \& Ostriker 2004; Madau et al. 2004; Shapiro 2005; Volonteri \& Rees 2005). The duration and the efficiency of  accretion are highly uncertain, making hard to determine the amount of accreted mass. Shapiro (2005) suggests that the IMBH mass evolution $m_{\bullet{}}(t)$, assuming  Eddington rate accretion, can be written as:
\q\label{eq:eq8}
m_{\bullet{}}(t)=m_{\bullet{},f}\,{} \exp{\left(\frac{1-\epsilon{}}{\epsilon}\,{}\frac{t}{t_{Salp}}\right)},
\nq
where  $\epsilon$ is the radiative efficiency ($\epsilon{}\approx{}0.1$) and $t_{Salp}$ is the Salpeter time ($t_{Salp}\approx{}0.45$ Gyr). In our case $t$ is the fraction of  IMBH life during which it accretes at the Eddington rate, i.e. $t=f_{duty}\,{}t_{birth}$, where $t_{birth}$ is the time elapsed from the IMBH formation ($\approx{}13.5$ Gyr) and $f_{duty}$ is the fraction of  $t_{birth}$ during which the IMBH accretes. Assuming  $f_{duty}\approx{}0.01$ ($f_{duty}\lesssim 0.03$ for quasars at $z\approx{}6$, Steidel et al. 2002), we obtain $m_{\bullet{}}(t)\approx{}10\,{}m_{\bullet{},f}$. For our choice of $m_{\bullet{},f}=10^3M_\odot$, this means that the average mass of IMBHs today is  $m_{\bullet{}}(t)\approx 10^4M_\odot$, consistent with the value $1.8\times{}10^4M_\odot$ of the recently detected IMBH candidate in the globular cluster G1 (Gebhardt et al. 2005) and with previous theoretical estimates (Volonteri et al. 2003). This estimate is affected by a number of uncertainties, and we consider it only as an educated guess.\\
Due to accretion, the current density of IMBHs $\Omega_{\bullet{}}$ will be 
\q\label{eq:eq9}
\Omega_{\bullet{}}=\frac{m_{\bullet{}}(t)}{m_{\bullet{},f}}\,{}\Omega_{\bullet{},\,{}f} \approx 10\,{}\Omega_{\bullet{},\,{}f} = 10^{-3}\Omega{}_b (m_{\bullet{},f}/10^3M_\odot),
\nq
our reference value. 

We note that other models predict very different values for $\Omega_{\bullet{}}$. For example, Salvaterra \& Ferrara (2003) derived $\Omega_{\bullet{}}\approx{}0.1\,{}\Omega_b$, under the assumption that Population III stars are the sources of the observed near-infrared excess with respect to galaxy counts.
One of the aims of this paper is to check which part of the $\Omega_{\bullet{}}$  range is allowed by the link between IMBHs and ULXs (see next section). For this reason, we also carried out two runs adopting the estimate by Salvaterra \& Ferrara (2003).

The last problem we have to address is the selection of initial conditions for the position and velocity distribution of IMBHs. White \& Springel (2000) suggested that remnants of Population III stars should be much rather concentrated inside present-day halos. N-body cosmological simulations by Diemand, Madau \& Moore (2005, hereafter DMM) seem to support this idea; they also show that the present spatial distribution of objects formed in high-$\sigma{}$ fluctuations depends only on the rarity of the peak in which they are born. In particular, DMM find that the spatial distribution, in present halos, of objects formed in a $\nu{}\,{}\sigma{}$ fluctuation is well fitted by a modified Navarro-Frenk-White profile:
\q\label{eq:eq10}
\rho{}_{\bullet{}}(r)=\frac{\rho_s}{(r/r_\nu)^\gamma{}\,{}(1+(r/r_\nu{})^\alpha{})^{(\beta{}_\nu{}-\gamma{})/\alpha{}}},
\nq
where $\rho_s$, $\alpha{}$ and $\gamma{}$ are the same as defined in the previous section;  $r_\nu\equiv{}r_s/f_\nu{}$ is the scale radius for objects formed in a  $\nu{}\,{}\sigma{}$ fluctuation (with $f_\nu{}=\exp{(\nu{}/2)}$), and $\beta{}_\nu{}=3+0.26\,{}\nu{}^{1.6}$.
As DMM simulations are collisionless, they cannot take into account the possible formation of binaries containing IMBHs (eventually with the central SMBH) and the occurrence of three-body encounters, which likely lead to the ejection of one of the involved IMBHs (Volonteri et al. 2003). Thus, the actual IMBH distribution could be slightly more ''diluted'' than that obtained by DMM.

%%%%%%%%%%%%%%%%%%%%%%%%%%%%%%%%%%% FIGURE 1 %%%%%%%%%%%%%%%%%%%%%%%%%%%%%%%%%%
\begin{figure}
\center{{
\epsfig{figure=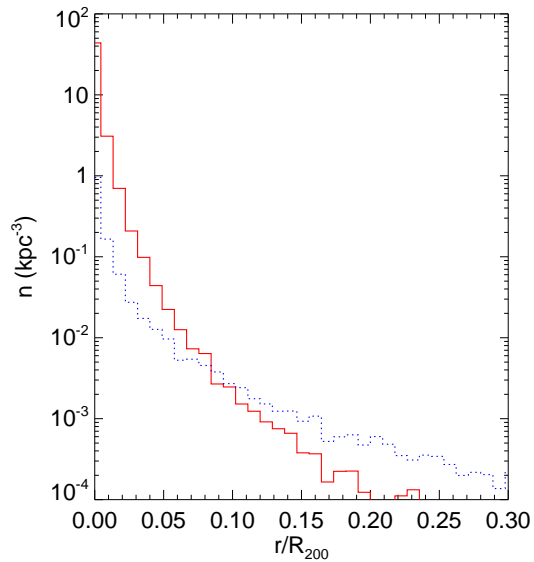,height=8cm}
}}
\caption{\label{fig:fig1} 
Density profile of IMBHs for a DMM (case A1;  solid line) and for a NFW distribution (case A2;  dotted line).
} 
\end{figure}
%%%%%%%%%%%%%%%%%%%%%%%%%%%%%%%%%%%%%%%%%%%%%%%%%%%%%%%%%%%%%%%%%%%%%%%%%%%%%%%
%%%%%%%%%%%%%%%%%%%%%%%%%%%%%%%%%%% FIGURE 2 %%%%%%%%%%%%%%%%%%%%%%%%%%%%%%%%%%
\begin{figure*}
\center{{
\epsfig{figure=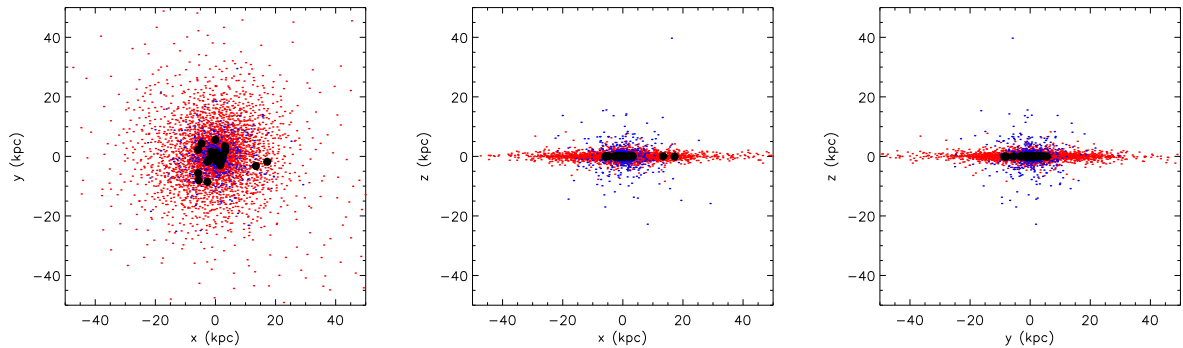,height=5cm}
}}
\caption{\label{fig:fig2} 
Snapshots at $t =1.4$~Gyr (about 5 Galactic timescales) of the simulated Milky Way; only 1/200 of the total bulge and 
disk particles are plotted.  Big dots indicate the IMBHs passing through the molecular disk in the case A1.  Left panel: particle positions in the $x-y$ plane;   central: in the $x-z$ plane;  right: in the $y-z$ plane.
} 
\end{figure*}
%%%%%%%%%%%%%%%%%%%%%%%%%%%%%%%%%%%%%%%%%%%%%%%%%%%%%%%%%%%%%%%%%%%%%%%%%%%%%%% 
\subsection{Description of runs}
We made three different sets of simulations, A, B and C, whose characteristics are described in Table 2. Simulations labeled as A are characterized by $\Omega_{\bullet{}}=10^{-3}\Omega_b$ (corresponding to IMBHs formed in 3 $\sigma{}$ fluctuations), simulations B have $\Omega_{\bullet{}}=10^{-1}\Omega_b$ (corresponding to the Salvaterra \& Ferrara 2003 model), and the simulation C has  $\Omega_{\bullet{}}=3\times{}10^{-5}\Omega_b$ (corresponding to IMBHs formed in 3.5 $\sigma{}$ fluctuations).
Simulations A1, B1 and C adopt the DMM spatial distribution, assuming that IMBHs form in 3 $\sigma$ (A1, B1) or 3.5 $\sigma{}$ (C) peaks. Instead, runs A2 and B2 were performed assuming that IMBHs follow a normal NFW profile. In Fig.~1 we compare the two considered distributions of IMBHs, i.e. DMM and NFW.
In each simulation we evolved about $10^6$ particles, each having a mass of $5\times{}10^4\,{}M_\odot{}$ (included the IMBH particles). These particles are divided in $\approx{}8\times{}10^5$ disk particles and $\approx{}2\times{}10^5$ bulge particles, plus a variable number of IMBH particles: 2000 in each run of the group A (corresponding to 10$^4$ IMBHs of $10^4\,{}M_\odot{}$; see equations (\ref{eq:eq7}) and (\ref{eq:eq8})), 2$\times{}10^5$ in each run of the group B (corresponding to 10$^6$  IMBHs of $10^4\,{}M_\odot{}$) and 100 in the run C (corresponding to 500  IMBHs of $10^4\,{}M_\odot{}$). CPU time limits require that we consider only equal mass particles, with mass no lower than  $5\times{}10^4\,{}M_\odot{}$ (included the IMBH particles), making impossible to investigate the impact of dynamical friction on the IMBH spatial distribution.
%%%%%%%%%%%%%%%%%%%%%%%%%%%%%%%%%%% TABLE 2 %%%%%%%%%%%%%%%%%%%%%%%%%%%%%%%%%%%
\begin{table}
\begin{center}
\caption{Initial parameters for the IMBHs.
}
\begin{tabular}{llll}
\hline
\hline
\vspace{0.1cm}
Run & Number of IMBH particles & $\Omega_{\bullet{}}/\Omega_b$ & IMBH profile\\
\hline
A1 & 2000 & 10$^{-3}$ & DMM$^{a}$\\
A2 & 2000 & 10$^{-3}$ & NFW$^{b}$\\
B1 & 2$\times{}10^5$ & 10$^{-1}$ & DMM\\
B2 & 2$\times{}10^5$ & 10$^{-1}$ & NFW\\
C & 100 & $3\times{}10^{-5}$ & DMM\\
\hline
\end{tabular}
\end{center}
\begin{flushleft}
{\footnotesize $^{a}$Diemand, Madau \& Moore 2005.}\\
{\footnotesize $^{b}$Navarro, Frenk \& White 1996.}\\
\end{flushleft}
\label{tab_2}
\end{table}
%%%%%%%%%%%%%%%%%%%%%%%%%%%%%%%%%%%%%%%%%%%%%%%%%%%%%%%%%%%%%%%%%%%%%%%%%%%%%%%

\section{IMBHs accreting molecular gas}
%One of the possible explanations for the existence of ULXs is that they are accreting IMBHs. Many works are present in literature, which  study the characteristics of accreting IMBHs in binary systems (Baumgardt et al. 2004;  Hopman \& Portegies Zwart 2005; Hopman, Portegies Zwart \& Alexander 2004; Kalogera et al. 2004; Patruno et al. 2005; Portegies Zwart, Dewi \& Maccarone 2004). However, very few optical counterparts of ULXs have been unmistakably detected and can be identified with companion stars (Liu et al. 2005). Then, it is still open the possibility that ULXs are IMBHs accreting gas during the transit through a dense molecular cloud, as recently suggested by Mii \& Totani (2005).
As discussed in the Introduction, one of the possible explanations for the existence of ULXs is that they are IMBHs, accreting both in binary systems (Patruno et al. 2005) or in molecular clouds (Mii \& Totani 2005). Here we investigate the possibility that ULXs are IMBHs accreting gas during their transit within a molecular cloud. We also checked how many non-ultra-luminous X-ray sources ($L_X<10^{39}$ erg s$^{-1}$) could be produced by IMBHs passing through molecular clouds.

\subsection{IMBH density constraints from ULXs}

The X-ray luminosity\footnote{More precisely, equation (11) refers to the bolometric luminosity due to the Bondi-Hoyle accretion. However, detailed models of spectra of black holes accreting in the Bondi-Hoyle regime (Beskin \& Karpov 2005) or forming ADAF disks (Narayan, Mahadevan \& Quataert 1998) show that more than 60\% of the bolometric luminosity is emitted in the X-ray range and more than 40\% between 0.2 and 10 keV (approximately the bandpass of Chandra and XMM). Because of the other uncertainties in our calculations, we think that the approximation that nearly all the Bondi-Hoyle luminosity is emitted in the X-ray band is acceptable.}
 of a BH with mass $M_{\bullet{}}$, expected from the Bondi-Hoyle accretion in a gas cloud, can be expressed as (Edgar 2004; Mii \& Totani 2005):
\q\label{eq:eq11}
L_X(\rho{},v)=4\,{}\pi{}\,{}\eta{}\,{}c^2G^2M_{\bullet{}}^2\rho{}\,{}\tilde{v}^{-3},
\nq
where $\eta{}$ takes into account the radiative efficiency and the uncertainties in the accretion rate, $c$ is the light speed, $\rho{}$ the density of the molecular cloud and $\tilde{v}=(v^2+\sigma{}^2_{MC}+c^2_s)^{1/2}$, $v$ being the IMBH velocity relative to the gas; $\sigma_{MC}$ and $c_s$ are the molecular cloud turbulent velocity and gas sound speed, respectively. Recently, Krumholz, McKee \& Klein (2006) have shown that  accretion rates in a turbulent medium might slightly differ from the above one. Because of the many uncertainties in our model, we do not attempt to deal with these subtleties. 
From  equation (\ref{eq:eq11}) and following Agol \& Kamionkowski (2002), Mii \& Totani (2005) derive the  number of ULXs with luminosity higher than $L_X$ as\footnote{We consider only the particular case of the equation by Mii \& Totani (2005) in which $M_{\bullet{}}\simgt{}10^3M_{\odot}$}:
\qa\label{eq:eq12}
%\begin{array}{l}
N_{ULX}(>L_X)\approx{}2.2\times{}10^{-2}N_{\bullet{}}\,{}f_{disk}\,{}\mu{}^{-1}\,{}\eta{}\hspace{1cm}\nonumber\\
\hspace{2.5cm}\times{}\left(\frac{M_{\bullet{}}}{10^4M_\odot}\right)^2 \,{}\left(\frac{10^{39}\textrm{erg s}^{-1}}{L_X}\right),
%\end{array}
\nqa
where $\mu{}$ is the mean molecular weight ($\mu{}\approx{}$1.2-2.3 depending on the fraction of H$_2$ molecules), $N_{\bullet{}}$ is the number of IMBHs in the Milky Way (see equation \ref{eq:eq7}) and $\eta{}$ is the radiative efficiency. The correct value of $\eta{}$ is completely uncertain. In fact, we do not even know whether an accretion disk forms at all. Agol \& Kamionkowski (2002) show that most of BHs accreting gas should form accretion disks; but these disks are not necessarily thin. If the accreting gas is able to form a thin accretion disk (Shakura \& Sunyaev 1974), then $\eta{}=0.1$, as assumed by Mii \& Totani (2005). However, it seems to be more realistic that the gas forms an ADAF (i.e. Advection-Dominated Accretion Flow) disk, whose radiative efficiency\footnote{The luminosity for an ADAF disk scales as $\dot{M}^2$, where $\dot{M}$ is the accretion rate. However, if log$(\dot{M}/\dot{M}_{Edd})\sim{}-4,-2$, where $\dot{M}_{Edd}$ is the Eddington accretion rate, the efficiency of an ADAF disk is about two orders of magnitude lower than the efficiency of a thin disk (see Figure 7 of Narayan et al. 1998). We can assume that the efficiency of an ADAF disk is $\eta{}=10^{-3}$, because the accretion rates of the IMBHs we are considering fall in the range above.} is of the order of $\eta{}=0.001$ for IMBHs of mass $M_\bullet{}\sim{}10^4\,{}M_\odot{}$ (Quataert \& Narayan 1999). Finally, models which take into  account gas magnetization (Beskin \& Karpov 2005) show that a high efficiency ($\eta{}\approx{}0.1$) is allowed, even if the thin disk does not form. To decide among these different models is beyond the scope of this paper; we will consider the two different values $\eta{}=0.1$ and $\eta{}=0.001$ bracketing the above possibilities in all our cases.

In equation (\ref{eq:eq12}) $f_{disk}$ is the fraction of IMBHs passing through the molecular disk of the Galaxy, for which we  assume a scale height $z_{MC}=75$ pc and a radial extension $R_{MC}\approx{}20$ kpc (Sanders, Solomon \& Scoville 1984). Mii \& Totani estimated $f_{disk}\approx{}4.5\times{}10^{-4}$, based on the hypothesis that IMBHs are a halo population following a standard NFW profile. Our simulations allow a more precise and direct estimate of  $f_{disk}$ from our simulations. As an example, in Fig. 2 we show a snapshot of our simulations, where the positions of IMBHs passing through the molecular disk are shown. Table 3 reports the simulated values for $f_{disk}$ and $N_{ULX}$ with luminosity $L_X \ge 10^{39}$~erg~s$^{-1}$.

For $\eta{}=0.1$, we find that, if IMBHs are born in 3 $\sigma{}$ fluctuations (corresponding to $\Omega{}_{\bullet{}}\approx{}10^{-3}\Omega_b$) and  their present distribution in the Milky Way follows the DMM model (case A1), the number of ULXs associated with these IMBHs is still consistent with zero ($N_{ULX}\approx{}0.2$; Table 3, third column). Instead, if $\Omega{}_{\bullet{}}\approx{}10^{-1}\Omega_b$ (case B1), the Milky Way should host about 30 active ULXs. Then, we conclude that, if the IMBHs  follow a DMM distribution, $\Omega{}_{\bullet{}}\approx{}10^{-3}\Omega_b$ can be considered as an upper limit for the present density of IMBHs. 
If, on the contrary, the IMBHs  follow a standard NFW profile, as assumed by Mii \& Totani (2005), the number of ULXs obtained for $\Omega{}_{\bullet{}}\approx{}10^{-1}\Omega_b$ (case B2) is still marginally consistent with zero. Finally, if IMBHs follow the DMM distribution but form only in fluctuations with $\sigma{}\simgt{}3.5$, they are so rare that no ULX is expected to be seen in the Milky Way (case C1). 

However, if IMBHs are surrounded by low efficiency ADAF disks ($\eta{}=10^{-3}$), the upper limit for a DMM profile becomes $\Omega{}_\bullet{}=0.1\,{}\Omega{}_b$; whereas there are nearly no constraints for the NFW profile. It is worth noting that 
%for ULXs Mii \& Totani (2005) only considered the case of maximal efficiency ($\eta{}=0.1$), which is probably unlikely according to Agol \& Kamionkowski (2002).
Mii \& Totani (2005) mainly considered the case of maximal efficiency ($\eta{}=0.1$), which is probably unlikely according to Agol \& Kamionkowski (2002).
%
%%%%%%%%%%%%%%%%%%%%%%%%%%%%%%%%%%% FIGURE 3 %%%%%%%%%%%%%%%%%%%%%%%%%%%%%%%%%%
\begin{figure*}
\center{{
\epsfig{figure=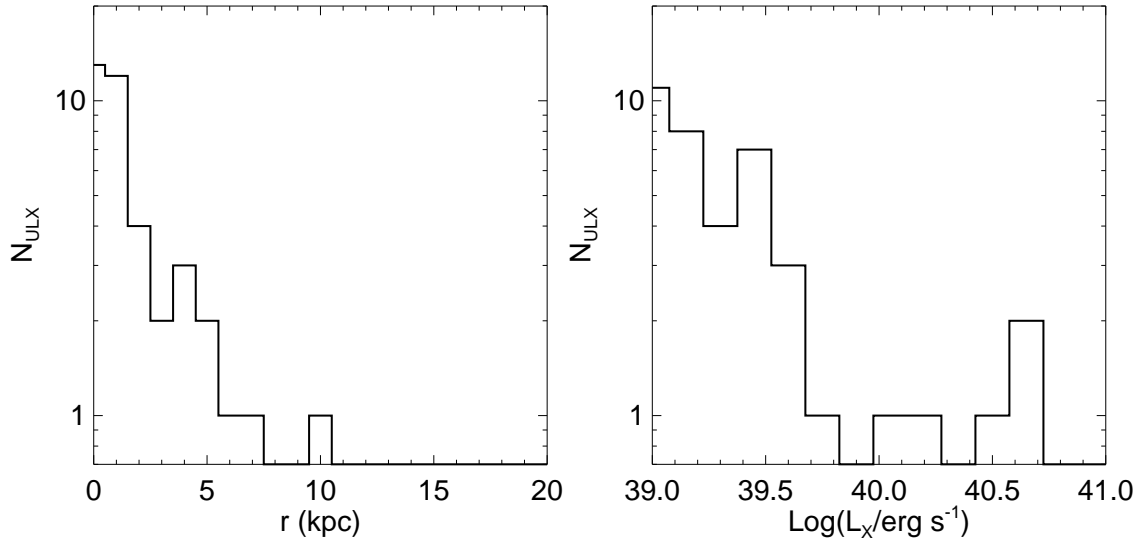,height=8cm}
}}
\caption{\label{fig:fig3} 
Distribution of ULXs as a function of their Galactocentric distance (left panel) and X-ray luminosity 
(right panel) for the case B1 and $\eta{}=0.1$, after $\approx 0.5 Gyr$.
} 
\end{figure*}
%%%%%%%%%%%%%%%%%%%%%%%%%%%%%%%%%%%%%%%%%%%%%%%%%%%%%%%%%%%%%%%%%%%%%%%%%%%%%%%

Equation \ref{eq:eq12} tells us even another information: the dark matter halo of the Milky Way cannot be entirely composed by BHs with mass $\gtrsim{}10^5\,{}M_\odot{}$. In fact, if $M_\bullet=10^5\,{}M_\odot{}$ and $N_\bullet{}=10^7$ (corresponding to the assumption that the Milky Way dark matter halo is composed by  BHs as massive as $10^5\,{}M_\odot{}$), $N_{ULX}(>10^{39}\textrm{erg s}^{-1})\sim{}2\times{}10^5\,{}\eta{}$ for a DMM profile and $N_{ULX}(>10^{39}\textrm{erg s}^{-1})\sim{}9\times{}10^3\,{}\eta{}$ for a NFW model. This means $N_{ULX}\gg{}1$ both for a thin and an ADAF disk model (unless $\eta{}\ll{}10^{-3}$). Then, the dark matter halo can be entirely composed by BHs only if their mass is less than $10^5\,{}M_\odot{}$, ruling out the scenario proposed by Lacey \& Ostriker (1985), in which halo BHs can account for the galactic disk heating.

\subsection{Radial and luminosity distribution of ULXs} 
The previous results can be refined by using our simulations instead of equation (\ref{eq:eq12}). In fact, from the simulations we know the number $N(z<z_{MC})$ of IMBH particles which at a given time are in the molecular disk (defined by the scale height $z_{MC}=75$ pc and the radial extension $R_{MC}\approx{}20$ kpc). It is well known that  H$_2$ in the Galaxy is not distributed uniformly within such disk, but it has a clumpy structure made of molecular clouds. Following Agol \& Kamionkowski (2002) we  derive the actual volume fraction of the molecular disk occupied by the clouds, i.e.
\qa\label{eq:eq13}
%\begin{array}{l}
f_{MC}=\frac{(\beta{}-2)\langle{}\Sigma{}_{MC}\rangle}{(\beta{}-1)2\,{}\mu{}\,{}m_p\,{}z_{MC}\,{}n_{min}}\left[1-\left(\frac{n_{max}}{n_{min}}\right)^{(1-\beta{})}\right]\hspace{-5.6cm}\nonumber\\
\approx{}0.017,
%\end{array}
\nqa
where $\beta{}=2.8$ for an H$_2$ cloud, $\langle{}\Sigma{}_{MC}\rangle=29\,{}M_\odot{}$ pc$^{-2}$ (Sanders et al. 1984; Mii \& Totani 2005) is the average surface density of molecular clouds,  $m_p$ is the proton mass,  $n_{min}=10^2$ cm$^{-3}$ and  $n_{max}=10^5$ cm$^{-3}$ are the minimum and maximum density, respectively, observed in molecular clouds. 
Thus, the number of IMBHs which at a given time $t$ are embedded into a molecular cloud  is $N(z<z_{MC})f_{MC}$. In practice, we randomly select from our simulations a fraction $f_{MC}$ of the IMBHs which at a given time $t$ have $z<z_{MC}$ and $R<R_{MC}$. For this sample of IMBHs, we derive the Bondi-Hoyle luminosity $L_X$  as in equation (\ref{eq:eq11}), assuming
that $c_s=0.3$ km s$^{-1}$ and $\sigma_{MC}=3.7$ km s$^{-1}$ (Larson 1981; Solomon et al. 1987; Mii \& Totani 2005). We then identify as ULXs those IMBHs which have $L_X>10^{39}$ erg s$^{-1}$. Averaging this number over the entire simulation, we obtain an estimate $\tilde{N}_{ULX}(>L_X)$ of the number of ULXs in the Milky Way.
% (Table 3, fourth column). 
For all the considered cases, the number $\tilde{N}_{ULX}(>L_X)$ (Table 3, fourth column), derived in this way, is consistent with the value $N_{ULX}(>L_X)$ (Table 3, third column), derived from equation (\ref{eq:eq12}),
confirming the validity of the Mii \& Totani calculation.

This alternative method to derive the number of ULXs contains  additional important pieces of  information concerning  the spatial distribution of ULXs and their luminosities (see Fig.~3 for the case B1). These distributions are meaningless for the Milky Way, where no ULXs have been detected. Nevertheless, it could be interesting to compare them with the distributions of ULXs observed in other spiral galaxies.
Fig.~3 shows that, if a DMM profile is adopted for IMBHs, ULXs appear to be mostly concentrated towards the Galactic center.  This seems to be at odds with observations, which have shown that ULXs of external galaxies tend to be preferentially located in spiral arms (Liu \& Bregman 2005). We have to keep in mind, though, that the present calculation is based on the molecular hydrogen distribution of the Milky Way, which could be quite different from that of other galaxies hosting ULXs; the latter are often starbursting, very gas rich systems. 

The predicted ULX  luminosities (Fig.~3) are mostly in the range 1 - 5 $\times{}10^{39}$ erg s$^{-1}$ with only few sources showing   luminosities higher than $10^{40}$ erg s$^{-1}$. This rapid falloff of the number of ULXs for increasing X-ray luminosities is consistent with observations (Grimm et al. 2003). On the contrary,  simulations following the accretion of IMBHs in binary systems indicate a number of low luminosity ULXs which is only a factor $\approx{}2$ higher than the number of sources with $L_X>10^{40}$ erg s$^{-1}$ (Madhusudhan et al. 2005). As a caveat, we recall that our calculations assume a Bondi-Hoyle luminosity, which might be a relatively oversimplified approximation.
\subsection{Non ULX sources}
By using the technique described in Section 3.2, we can also derive
the number of low luminosity X-ray (in brief, non-ULX) sources with
$L_X < 10^{39}$ erg s$^{-1}$ (Table 3; fifth column).  An interesting
result is that, if $\eta{}=0.1$, IMBH luminosities are always as high
as $10^{37}$ erg s$^{-1}$ (see Fig.~4, where the dotted line
represents IMBHs accreting molecular gas, including ULXs). Instead, if only ADAF disks can form, the luminosities
reached by accreting IMBHs in molecular clouds are lower, in the range
from $10^{35}$ to $10^{38}$~erg s$^{-1}$. Luminosities from $10^{37}$ to
$10^{39}$ erg s$^{-1}$ are reached, in our Galaxy, only by a few
supernova remnants (Vink 2006) and by high mass and low mass X-ray
binaries (HMXBs, LMXBs; Psaltis 2006). Then, IMBHs accreting gas in molecular
clouds should be among the most powerful Galactic X-ray sources and
therefore should have been already detected, provided they are not
transient.  A strong constraint on the density of IMBHs thus descends
from the requirement that the number of IMBHs with $L_X<10^{39}$ erg
s$^{-1}$ is lower than the number of detected Galactic sources
emitting at the same luminosities which lack of certain
identifications with other kind of objects (such as HMXBs and
LMXBs). This analysis will be carried out in Section 5, considering
X-ray sources produced by IMBHs accreting both within molecular clouds
and atomic hydrogen regions.

\begin{table*}
\begin{center}
\caption{Results.}
\begin{tabular}{lllllll}
\hline
\hline
\vspace{0.1cm}
Run & $f_{disk}$$^{a}$ & $N_{ULX}$$^{b}$ & $\tilde{N}_{ULX}$$^{c}$ & $N_{H_2}$$^{d}$ & $N_{H}$$^{e}$ & $N(10^{36-39}$ erg s$^{-1})$$^{f}$\\% & $N_{tid}$$^{d}$\\
\hline
%\multicolumn{6}{c}{$\eta{}=0.1$}\\
%\hline
A1 & 0.022$\pm{}$0.003 & 0.2 (0.002) & 0.2$\pm$0.2 (0.002$\pm$0.002) & 1.0$\pm{}$0.6 (1.2$\pm{}$0.5) & 45$\pm{}$12 (45$\pm{}$12) &  18$\pm{}$7 (1.2$\pm{}$1.0) \\%4$\pm{}$3 (0.23$\pm{}$0.23) \\
A2 & 0.0008$\pm{}$0.0006 &0.008 (8$\times{}10^{-5}$) & 0$\pm{}$0 (0$\pm{}$0) &  0$\pm{}$0 (0$\pm{}$0) & 5$\pm{}$4 (5$\pm{}$4) & 0.4$\pm$0.4 (0$\pm$0)\\%0.04$\pm$0.04 (0$\pm$0)\\
B1 & 0.025$\pm{}$0.002 & 28 (0.28) & 40$\pm{}$6 (0.5$\pm{}$0.5) & 310$\pm{}$24 (350$\pm{}$24) & 4056$\pm{}$125 (4058$\pm{}$125) & 1650$\pm{}$70 (236$\pm{}$15)\\%527$\pm{}$37 (41$\pm{}$7)\\
B2 & 0.00090$\pm{}$0.00007 & 1 (0.01) & 0.5$\pm{}$0.5 (0.007$\pm{}$0.007) & 5.6$\pm{}$1.5 (6.1$\pm{}$1.5) & 495$\pm{}$39 (495$\pm{}$39) &  148$\pm{}$21 (5$\pm{}$3) \\%27$\pm{}$9 (0.5$\pm{}$0.5) \\
C &  0.039$\pm{}$0.018 & 0.02 (0.0002) & 0$\pm{}$0 (0$\pm{}$0) &  0$\pm{}$0 (0$\pm{}$0) & 3$\pm{}$3 (3$\pm{}$3) & 0.09$\pm{}$0.09 (0$\pm{}$0)\\ %0.001$\pm{}$0.001 (0$\pm{}$0)\\
\hline
\end{tabular}
\end{center}
\begin{flushleft}
{\footnotesize The values refer to a thin disk with $\eta=0.1$  (the values in parenthesis refer to an ADAF disk with $\eta=0.001$).}\\
{\footnotesize $^{a}$Average fraction of IMBHs passing through the molecular disk (see Section 3.1).}\\
{\footnotesize $^{b}$Number of ULXs with $L_X\ge 10^{39}$ erg s$^{-1}$ derived from equation (12) adopting $\mu{}=2$  (see Section 3.1).}\\
{\footnotesize $^{c}$Number of ULXs with $L_X\ge 10^{39}$erg s$^{-1}$ derived from our simulations (see Section 3.2 and Fig.~3).}\\
{\footnotesize $^{d}$Number of sources which accrete molecular hydrogen (see Section 3.3) with  X-ray luminosities $L_X<10^{39}$ erg s$^{-1}$, derived from our simulations (see Fig.~4 and 5).}\\
{\footnotesize $^{e}$Number of sources which accrete atomic hydrogen (see Section 4), derived from our simulations. All of them have $L_X\leq10^{39}$ erg s$^{-1}$ (see Fig.~4 and 5).}\\
{\footnotesize $^{f}$Number of sources which accrete atomic or molecular hydrogen and have X-ray luminosity $10^{36}\leq{}L_X<10^{39}$ erg s$^{-1}$  (see Section 5).}\\
%{\footnotesize $^{d}$Number of IMBHs which, during 11Gyr, pass within 0.01 kpc from the galactic center.}\\
\end{flushleft}
\label{tab_3}
\end{table*}
%%%%%%%%%%%%%%%%%%%%%%%%%%%%%%%%%%%%%%%%%%%%%%%%%%%%%%%%%%%%%%%%%%%%%%%%%%%%%%%

%%%%%%%%%%%%%%%%%%%%%%%%%%%%%%%%%%% FIGURE 4 %%%%%%%%%%%%%%%%%%%%%%%%%%%%%%%%%%
\begin{figure*}
\center{{
\epsfig{figure=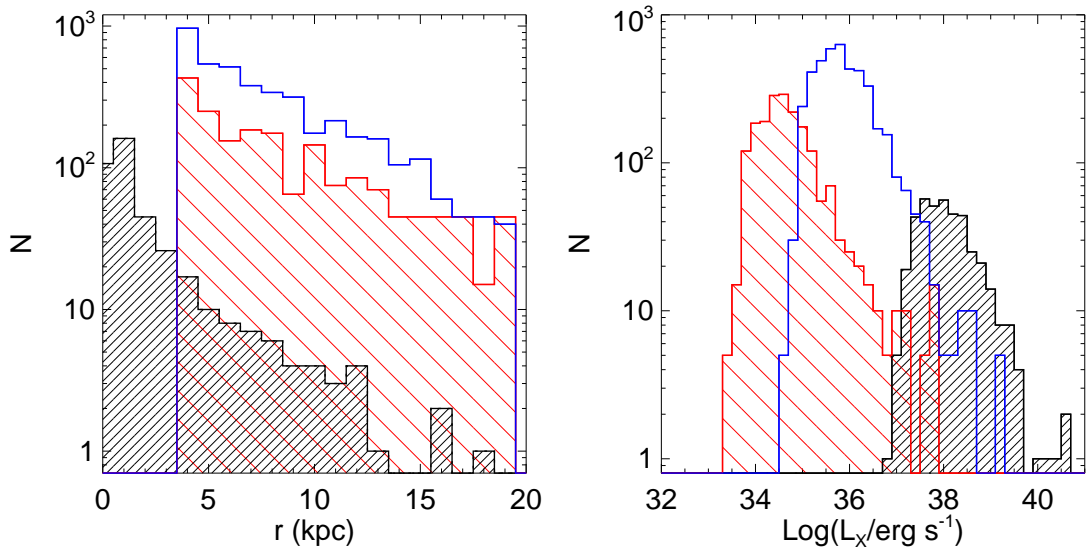,height=8cm}
}}
\caption{\label{fig:fig4} 
Distribution of all the X-ray sources as a function of the radial distance (left panel) and luminosity (right panel) for the case B1 and $\eta{}=0.1$, after 2 timescales. 
%Solid line:  IMBHs passing through cold neutral hydrogen;  dot-dashed: IMBHs passing through warm hydrogen,   dotted: IMBHs passing thorough molecular clouds.
Open histogram: IMBHs passing through cold neutral hydrogen; light shaded histogram: IMBHs passing through warm hydrogen; heavy shaded histogram: IMBHs passing through molecular hydrogen. Although the distributions slightly change with time due to the small statistics, their main features remain unaltered.
} 
\end{figure*}
%%%%%%%%%%%%%%%%%%%%%%%%%%%%%%%%%%%%%%%%%%%%%%%%%%%%%%%%%%%%%%%%%%%%%%%%%%%%%%%

%%%%%%%%%%%%%%%%%%%%%%%%%%%%%%%%%%% FIGURE 5 %%%%%%%%%%%%%%%%%%%%%%%%%%%%%%%%%%
\begin{figure*}
\center{{
\epsfig{figure=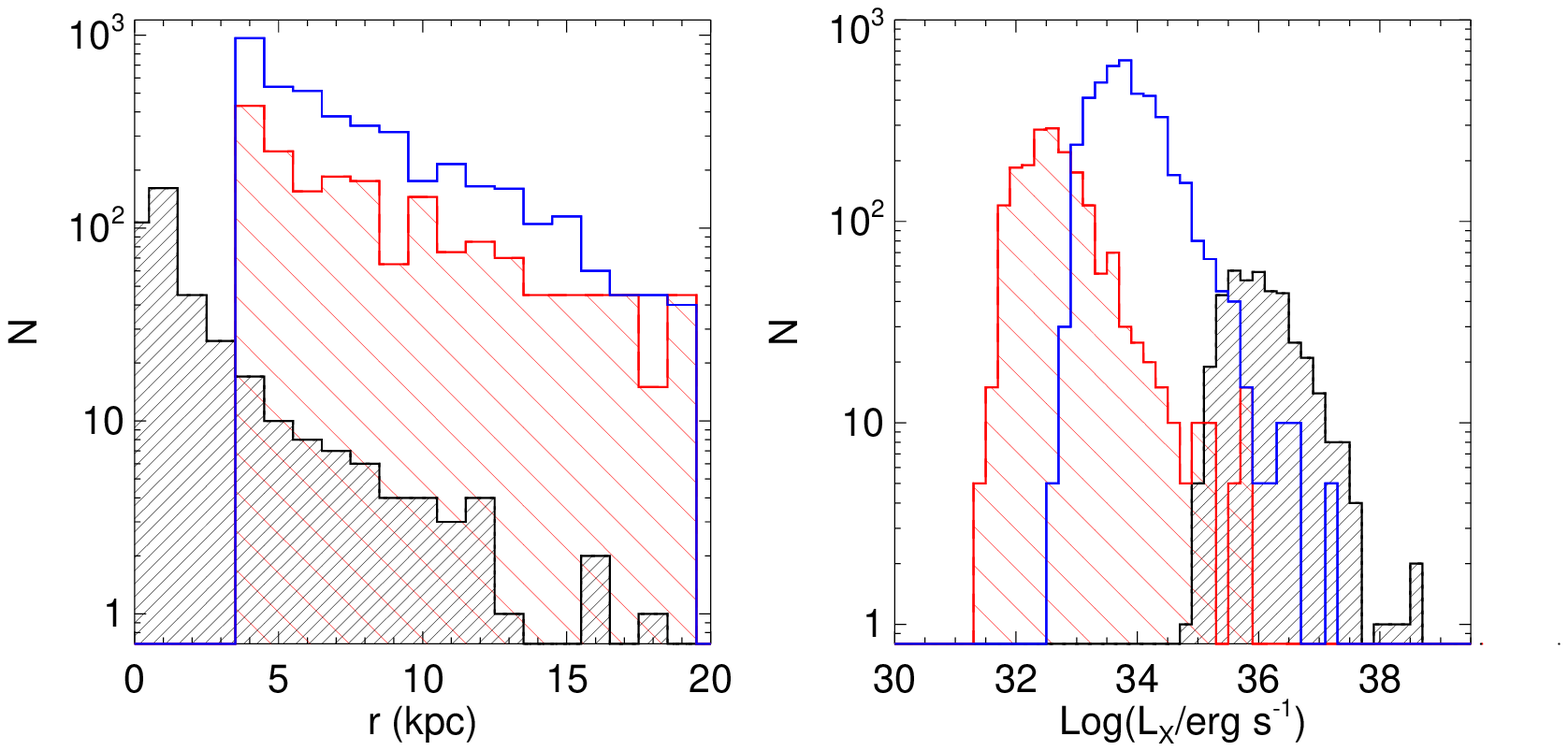,height=8cm}
}}
\caption{\label{fig:fig5} 
Distribution of all the X-ray sources as a function of their Galactocentric distance (left panel) and luminosity 
(right panel) for the case B1 and $\eta{}=0.001$, after $\approx 0.5$~Gyr. 
% Solid line:  IMBHs passing through cold neutral hydrogen;  dot-dashed: IMBHs passing through warm hydrogen,  dotted: IMBHs passing thorough molecular clouds.
Open histogram: IMBHs passing through cold neutral hydrogen; light shaded histogram: IMBHs passing through warm hydrogen; heavy shaded histogram: IMBHs passing through molecular hydrogen.
 Although the distributions slightly change with time due to the small statistics, their main features remain unaltered.
} 
\end{figure*}
\section{IMBHs accreting atomic hydrogen}
Mii \& Totani (2005) neglected in their analysis IMBHs passing through atomic hydrogen regions, because their lower density ($\lesssim{}1$ cm$^{-3}$) powers much lower X-ray luminosities than in molecular gas.
%. Accretion in regions less dense than molecular clouds is unlikely, because the emitted X-ray luminosity would be too low 
(King et al. 2001). However,  atomic hydrogen is much more diffuse in the Milky Way than H$_2$, and IMBHs are so massive that they can have non-negligible luminosity even accreting in such rarefied environment.
Current models of the hydrogen distribution in the Milky Way (McKee \& Ostriker 1977; Rosen \& Bregman 1995) suggest the existence of three different phases: a neutral cold component ($T\approx{}10^2$ K), a warm ($T\approx{}10^4$ K) and a hot ($T\approx{}10^6$ K) component. In our work we neglect the hot component, since, even if its filling factor is high (up to 0.7, Rosen \& Bregman 1995), it has an average density  of  $\approx{}10^{-3}$ cm$^{-3}$ and a sound speed of $\approx{}100$ km s$^{-1}$, so that the X-ray luminosity of IMBHs accreting hot gas is expected to be very low.
We define an atomic hydrogen disk as a disk having cut-off length $R_{H}=20$ kpc (the data show an exponential fall of neutral hydrogen density outside 20 kpc; Lockman 2002) and scale height $z_{H}=100$ pc (Baker \& Burton 1975; Sanders et al. 1984; Dickey \& Lockman 1990).   Adopting the procedure described in Agol and Kamionkowski (2002),
%tried to study the emission of IMBHs passing through neutral hydrogen regions with the same procedure as described in Section 3.2. 
%Instead of $f_{MC}$, we have to 
we derive the volume fraction occupied by cold neutral hydrogen, $f_{CH}$, as:
\qa\label{eq:eq14}
f_{CH}=\frac{(\beta{}_{CH}-2)\langle{}\Sigma{}_{CH}\rangle}{(\beta{}_{CH}-1)2\,{}\mu{}\,{}m_p\,{}z_{H}\,{}n_{min,CH}}\nonumber\\
\times{}\left[1-\left(\frac{n_{max,CH}}{n_{min,CH}}\right)^{(1-\beta{}_{CH})}\right],
\nqa
where $\beta{}_{CH}=3.8$ (Agol \& Kamionkowski 2002),  $\langle{}\Sigma{}_{CH}\rangle{}$ is the average surface density of neutral hydrogen ($\langle{}\Sigma{}_{CH}\rangle{}=4.5\,{}M_\odot{}$ pc$^{-2}$ if $R>4$ kpc and $\langle{}\Sigma{}_{CH}\rangle\approx{}0$ if $R\leq{}4$ kpc; Agol \& Kamionkowsky 2002; Sanders et al. 1984), $n_{min,CH}$ and $n_{max,CH}$ are the minimum and maximum density of neutral hydrogen, respectively ($n_{min,CH}\approx{}1$ cm$^{-3}$, $n_{max,CH}\approx{}5$ cm$^{-3}$; Bregman, Kelson \& Ashe 1993). Substituting these values into equation (\ref{eq:eq14}), we obtain $f_{CH}=0.48$ if $R>4$ kpc and $f_{CH}\approx{}0$ if $R\leq{}4$ kpc. This value is in good agreement with run E of Rosen \& Bregman (1995), which is a suitable fit of cold and warm hydrogen observations (Dickey \& Lockman 1990).  
For consistency, we assume that the filling factor of the warm component is $f_{WH}$=0.2, as  in run E of Rosen \& Bregman  (1995). 
Then, the number of IMBHs which, at a given time $t$, are passing through cold (warm) hydrogen regions is $N(z<z_{H})\,{}f_{CH}$ ($N(z<z_{H})\,{}f_{WH}$). Adopting the same  procedure as in Section 3.2, we randomly select a fraction $f_{CH}$ (for cold hydrogen) or $f_{WH}$ (for warm hydrogen) of the IMBHs which, at a given time, pass through the neutral hydrogen disk, and we derive the Bondi-Hoyle luminosity\footnote{We assume a turbulent velocity $\sigma{}_H=10$ km s$^{-1}$ both for cold and warm hydrogen (Lockman \& Gehman 1991). The adopted sound speed is 1 km s$^{-1}$ for cold hydrogen and 10 km s$^{-1}$ for warm hydrogen regions.} for each of them using equation (\ref{eq:eq11}). The results are shown in Table 3, sixth column, and in Figure 4. 

If $\eta{}=0.1$, IMBHs  passing through cold neutral hydrogen regions show  luminosities of the order of $10^{35-37}$~erg s$^{-1}$, with a high luminosity tail at $ >10^{38}$ erg s$^{-1}$; IMBHs passing through warm hydrogen regions produce lower luminosities, ranking from  $10^{33}$ to $10^{35}$ erg s$^{-1}$ with an extended high luminosity tail extending to 10$^{38}$ erg s$^{-1}$ (Fig.~4). Due to the large filling factor of the atomic hydrogen with respect to the molecular one, the number of IMBHs  accreting HI is a factor $\approx{}10-30$ higher than the number of IMBHs which accrete H$_2$, even if the luminosities are lower and nearly no ULX can be produced in atomic  regions.

Most interestingly,  we note from the column 6 of Table 3  that for $\Omega{}_\bullet{}=0.1\,{}\Omega_b$ (both for $\eta{}=0.1$ and for $\eta{}=0.001$), the expected number of X-ray sources  is huge, both for in the case of a DMM profile (case B1, $\approx{}4000$ sources) and for a NFW profile (case B2,  $\approx{}500$ sources). The reason the case B2 (which yielded the acceptable number of ULXs $\approx1$) predicts so many X-ray sources depends
on the HI distribution properties: atomic gas is less concentrated than molecular clouds. Because a number of X-ray sources (not identified with HMXBs or LMXBs) $> 500$ is definitely too high for the Milky Way, we can robustly exclude  $\Omega{}_\bullet{}\gtrsim{}0.1\,{}\Omega_b$.

For $\eta{}=0.001$, such as for an ADAF disk, IMBH luminosities are much lower, spanning  $10^{31-35}$ to $10^{37}$ erg s$^{-1}$ (Fig.~5). Even if the total number of sources remains nearly unmodified (Table 3; sixth column), the fact that most of them present luminosities  $ \ll 10^{37}$ erg s$^{-1}$ makes comparison with observations more difficult.
%Therefore, we can still reject the cases with $\Omega{}_\bullet{}\gtrsim{}0.1\Omega_b$ (B1 and B2), because the number of X-ray sources is far too large.
In the next section we attempt to constrain the Galactic IMBH density by comparison with X-ray observations, considering  X-ray sources produced collectively by IMBHs accreting both within molecular clouds and atomic  regions.
\section{Comparison with observations}
In the previous Sections (3.1-3.2) we tried to constrain the density
of IMBHs in our Galaxy by the fact that no ULXs have been detected 
in the Milky Way. From our simulations we found that some accreting
IMBHs might also emit as non-ultraluminous X-ray sources, being in
some cases bright enough to had been reliably detected by current X-ray satellites.

Hereafter, we match the results of our simulations with the X-ray
observations. In particular, we compare the predicted IMBH X-ray
emission with our knowledge of the X-ray sky in order to define an
upper limit on the presence of these objects in our Galaxy.  

So far we have predicted IMBH X-ray luminosities (for a summary see Table 3) of  $\approx10^{31-39}$\,erg\,s$^{-1}$, mostly depending on the assumed accretion efficiency $\eta$, disk model and molecular or atomic accreting material. Searching in the observations
for an upper limit of possible IMBHs in such a wide luminosity range is a non-sense, mainly because at the low luminosities many sources were certainly missed.

What we then study here are only the IMBHs with a predicted hard X-ray luminosity between $10^{36}-10^{39}$\,erg\,s$^{-1}$ (see Table\,3, last column).  In all these cases, the high luminosity of these sources makes us confident that we should have seen them in the monitoring campaign of the Galaxy with the new generation satellites (within a certain distance depending of the flux resolution of the given satellite).

The most uncertain point is whether IMBHs accreting gas are transient or persistent sources. If the IMBH would be able to 
form a (thin or ADAF) accretion disk,  it should also be a transient source. Instead, IMBHs accreting in the Bondi-Hoyle 
regime without forming a disk, as suggested by Beskin \& Karpov (2005), should show flares; it remains unclear if  they 
can be transient sources or not. On the other hand, the IMBH could be transient also as a consequence of properties of the 
interstellar medium. In fact the accretion rate is roughly proportional to the density of the gas, and the scintillation 
measures show that the density fluctuations of the interstellar medium can be as high as a factor 100 on scales from 
$\approx{}$10$^{18}$ down to $\approx{}$10$^{12}$ cm (Rickett 1990; Lambert \& Rickett 2000; Cordes \& Lazio 2001; 
Ferrara \& Perna 2001). A halo IMBH can easily travel $\approx{}$10$^{12}$ cm in about one day, and thus could suffer, 
in principle, changes of a factor $\approx{}$100 in its flux in this range of time. As a consequence, we have considered all the sources meeting our
requirements, both persistent or transient during the observations. 
%Obviously, we cannot reliably say whether the persistent sources detected by  IBIS/ISGRI are transients on timescales larger than the exposure time, but for our aims of inferring an upper limit that is not crucial.

The most wide and sensitive survey available for our aims is the soft gamma-ray survey recently obtained by the IBIS/ISGRI (Imager on Board INTEGRAL Satellite/INTEGRAL Soft Gamma-Ray Imager) instrument on board of the INTEGRAL satellite (Bird et al. 2004, 2006). This survey observed 50\% of the sky with a flux limit of 1\,mCrab in the 20--100\,keV energy range. 

Among more than 200 sources detected by the IBIS/ISGRI soft gamma-ray  survey scan, we excluded all the sources that certainly could not belong to the sample of possible IMBHs. In particular, we excluded all the well established X-ray binaries, which are a well known highly luminous Galactic class (both as transient and persistent sources). Furthermore we withdraw from our sample all the X-ray binaries  known to host a neutron star (e.g. either because showing pulsations or thermonuclear bursts). We then filtered for a
couple of highly energetic supernova remnants. After this first filtering we end up with a few tens of unknown objects.

Given the fact that what IBIS/ISGRI measures is a certain flux at Earth and not a luminosity, which is usually hard to derive because of the poorly known distances, we put the sample of sources we derived after the latter filtering, at distances between 1--15\,kpc\,, and we took all the sources with an inferred luminosity $10^{36}-10^{39}$\,erg\,s$^{-1}$, which implies in terms of flux all the uncatalogued IBIS/ISGRI objects with a detected flux (within their errors) $>$4.8 mCrab in the 20--40\,keV energy range. Note that this flux limit is derived placing a source emitting $10^{36}$\,erg\,s$^{-1}$ at 15\,kpc (e.g. the edge of our Galaxy), it would be detected by IBIS/ISGRI at a flux of 4.8 mCrab in the 20--40\,keV energy range, well above the flux limit of the survey. Hence we are confident that, if present, our putative Galactic IMBHs would had been detected in the 50\% of the Galaxy covered by the IBIS/ISGRI survey.  Note that the flux limit of 4.8 mCrab we assumed includes, for the completeness of our analysis, the worst case of the faintest source at the largest distance: the fact that we are looking for an upper limit on the number of these possible IMBHs allow us to make this assumption.\\

Under these assumptions, we found only 3 IBIS/ISGRI unidentified sources which match our requirements. These sources 
were all persistent during the IBIS observations. Their luminosity falls in the 10$^{36}$-10$^{39}$\,erg s$^{-1}$ range, all of them close to the 10$^{36}$\,erg s$^{-1}$ bound.  As the IBIS/ISGRI catalogue covers 50\% of the Galaxy, we then 
tentatively predict an upper limit of 6 sources with these characteristics in the entire Galaxy, if the volume observed 
is a fair sample.  In a few years all the Galaxy will be covered by the IBIS/ISGRI survey and our tentative extrapolation 
may be refined.

Let us now compare this number with that of IMBHs predicted by our simulations in the same luminosity range and reported in the last column of Table 3.

\centerline {\it a) Thin disks}

For a thin disk, even case A1 ($\Omega_\bullet=0.001\,{}\Omega_b$, DMM profile) yields $\approx{}18$ sources, a value three times higher than observed. Furthermore, 4 of these simulated sources have $L_X > 10^{37}$ erg
s$^{-1}$. From an additional run with  $\Omega_\bullet{}=10^{-4}\Omega_b$ and the DMM profile (not reported in Table 3 for simplicity) we saw that the number of sources with 10$^{36}<L_X<10^{39}$ erg s$^{-1}$ is 0.6$\pm{}$0.6. We conclude that the upper limit in the case of a Shakura-Sunyaev disk and a DMM profile is $\Omega_\bullet=10^{-4}-10^{-3}\Omega_b$, similar to the upper limit found by the number of ULXs alone (see Section 3.1-3.2). 
Instead, for a NFW profile the allowed density of IMBHs is $> \Omega_\bullet=10^{-3}\Omega_b$ (case A2; corresponding to 0.4$\pm{}$0.4 sources),  but definitely $ < \Omega_\bullet=10^{-1}\Omega_b$ (case B2; 148$\pm{}$21 expected sources), strengthening the
constraint we found from the number of ULX.

\centerline {\it b) ADAF disks}

For the more realistic case of an ADAF disk, the constraints we obtain from the comparison with the IBIS/ISGRI sources are stronger than for the number of ULXs alone. In fact, if we assume a DDM profile, the upper limit for the density of IMBHs is about
$\Omega_\bullet=10^{-3}\Omega_b$ (case A1; 1.2$\pm{}$1.0 expected sources), much lower than
$\Omega_\bullet=10^{-1}\Omega_b$ (case B1; 236$\pm{}$15 expected sources), derived from the number of ULXs. If we
consider a NFW model, the upper limit is $\Omega_\bullet=10^{-1}\Omega_b$ (case B2; 5$\pm{}$3 expected sources); 
whereas there were no significant constraints from the ULXs. In summary, we must take with care the results of this comparison between simulated and observed X-ray sources, because of the huge uncertainties of our model. However, from the comparison with the IBIS/ISGRI unidentified sources we derive, in general, much stronger constraints than from the number of ULXs. 

\section{Conclusions}     
In this paper we have simulated the dynamical and emission properties of putative IMBHs which could inhabit our Galaxy. IMBHs are modeled as a halo population, distributed following a NFW or a more concentrated DMM profile. We assumed that IMBHs, 
passing through molecular or atomic hydrogen regions, could accrete gas, forming X-ray sources (either ultra-luminous or 
not). From the comparison of our simulations with the number of ULXs in the Galaxy (Section 3.1-3.2) and with the 
non-ultraluminous unidentified X-ray sources in the IBIS/ISGRI catalogue, we have derived the most stringent 
(to our knowledge) upper limits on the density $\Omega_\bullet$ of IMBHs. The main results can be summarized as follows:
\begin{itemize}
\item{} If IMBHs accrete with efficiency $\eta{}=0.001$ (i.e. via an ADAF disk), we obtain a strong upper limit 
$\Omega_\bullet{}\le 10^{-2}\Omega_b$ for a DMM profile and $\Omega_\bullet{}=10^{-1}\Omega_b$ for a NFW profile.
\item{} If the IMBHs accrete with efficiency $\eta{}=0.1$ (i.e. if a thin accretion disk around the IMBH is formed), the upper limit of the density of IMBHs is  $\Omega_\bullet{}=10^{-3}\Omega_b$ for a DMM profile and $\Omega_\bullet{}=10^{-2}\Omega_b$ for a NFW profile.
\end{itemize}
These results are still affected by some model uncertainties,  as the emission mechanism and the IMBH distribution. 
In addition, computational requirements have forced us to use high and equal mass ($m_\bullet{}=5\times{}10^4\,{}M_\odot{}$) IMBH particles. 
%thus preventing us to study additional effects related to the dynamical friction. 
Constraints for lower IMBHs masses are expected to be weaker. We can guess how the above upper limits change for different 
values of the IMBH mass by using the equation (\ref{eq:eq12}). For example, if we assume $m_\bullet{}=10^3M_\odot{}$,  $\eta{}=0.1$ and a DMM profile, the upper limit of the IMBH density becomes $\Omega_\bullet{}=10^{-2}\Omega_b$, about one 
order of magnitude lower than for $m_\bullet{}=10^4M_\odot{}$. As a further caveat, this extension to lower masses is 
possible only for the comparison with ULXs (and not with IBIS/ISGRI sources), because it is based on eq.  
(\ref{eq:eq12}).  Therefore, higher resolution simulations would be required to extend our studies to lower mass IMBHs 
or to consider a more realistic IMBH mass spectrum. Higher resolution simulations (where the mass of star particles can be orders of magnitude lower than the mass of IMBH particles)  are also needed to take into account the dynamical friction, which could play a crucial role. 

Another caveat concerns the validity of the DMM profile. The simulations by DMM  neglect the contribution of IMBHs in building up SMBHs, either by mergers  (Islam, Taylor, \& Silk 2003, 2004a,b,c) or by accretion and close dynamical encounters (Volonteri et al. 2003). Monte Carlo simulations combined with semi-analytical models (Volonteri \& Perna 2005) show that, if all these factors are taken into account, the number of IMBHs could be up to 2 orders of magnitude lower, leading to an $\Omega{}_\bullet{}\sim{}10-100$ lower than our estimates, and therefore compatible with the non-detection of ULXs in the Milky Way. However DMM take into account the bias in the formation sites of IMBHs, the accretion into larger halos, the role of both dynamical friction and tidal stripping, which were neglected or described by rough models in the previous studies. Unfortunately current simulations cannot account for all these effects at the same time.
In conclusions, even if our results could be improved under many aspects, we consider it as a success 
that our models strengthen by a  factor 10-1000 the currently adopted upper limits for the density of IMBHs 
(i.e. $\Omega_\bullet{}\approx{}0.02$; van der Marel 2004).

\section*{Acknowledgements}
We thank E.~Ripamonti, L.~Mayer, A. Possenti, M.~Colpi, T.~Maccarone, T.~Di~Salvo, L.~Giordano, S.~Callegari and C.~Vignali for useful discussions and we acknowledge the Referee, Marta Volonteri, for her critical reading the manuscript. NR thanks R.~Turolla and P.~Jonker for useful discussion. We also thank A.~Bazzano, L.~Bassani e L.~Kuiper for informations about the IBIS/ISGRI survey. We acknowledge for technical support the staff at {\it Cilea} and the system managers of the Kapteyn Institute of Groningen.

%Formato per le figure
%\begin{figure}
%\center{{
%\epsfig{figure=name.ps,height=8cm}
%}}
%\caption{\label{name} caption}
%\end{figure}

\onecolumn
\appendix

\end{document}